# Large Bandgap Observed on the Surfaces of EuZn$_2$As$_2$ Single Crystals

Dejia Kong[1,2], Siavash Karbasizadeh[3], Ganesh Narasimha[1], Paras Regmi[3], Chenggang Tao[4], Sai Mu[3], Rama Vasudevan[1], Ian Harrison[2], Rongying Jin[3*], and Zheng Gai[1*]

[1] Center for Nanophase Materials and Sciences, Oak Ridge National Laboratory, Oak Ridge, TN 37831

[2] Department of Chemistry, University of Virginia, Charlottesville, VA 22903

[3] SmartState Center for Experimental Nanoscale Physics, Department of Physics and Astronomy, University of South Carolina, Columbia, SC 29208

[4] Institute for Critical Technology and Applied Science, Department of Materials Science and Engineering, Virginia Tech, Blacksburg, Virginia 24061

[*] Corresponding authors: gaiz@ornl.gov, rjin@mailbox.sc.edu

**Abstract**

EuM$_2$As$_2$ (M = Zn, Cd, In, Sn etc.) is an excellent material system for studying topological properties, which can be easily tuned by magnetism involved. Theoretical calculations predict gapped and flat bands in EuZn$_2$As$_2$ but gapless structure in EuCd$_2$As$_2$. In this work, low-temperature (77 K) cleaved EuZn$_2$As$_2$ crystals are studied using scanning tunneling microscopy/spectroscopy (STM/S) and density functional theory (DFT) calculations. Defects-induced local density of states (LDOS) modification with a triangular shape helps identify the surface terminations: Eu versus AsZn surface. While large bandgaps (~1.5 eV at 77 K) are observed on both pristine surfaces, the bandgap width is found to be very sensitive to local heterogeneity, such as defects and step edges, with the tendency of reduction. Combining experimental data with DFT simulations, we conclude that the modified bandgap in the heterogeneous area arises from Zn vacancies and/or substitution by As atoms. Our investigation offers important information for reevaluating the electron topology of the EuM$_2$As$_2$ family.

**Introduction**

In the search for magnetic topological materials, EuM$_2$As$_2$ (M = Zn, Cd, In, Sn, etc.) compounds have received extensive attention in the materials science community. The excitement was initially generated by intensive experimental and theoretical investigations on EuCd$_2$As$_2$, which suggested that its electronic structure is topologically protected because of its combined crystallographic and magnetic structures.[1,2] In particular, angle-resolved photoemission spectroscopy (ARPES) measurements indicate the topological phase transition from the Weyl state with a single pair of Weyl points above the Eu ordering temperature $T_N$ ~ 9 K to the Dirac state below $T_N$.[1,3] Eu moment direction in EuIn$_2$As$_2$ determines the formation of either the anion-insulating state or topological-insulating state.[4,5] Eu-based compounds are thus considered an ideal platform for studying the relationship between electronic topology and magnetism, with switchable topological states solely controlled by magnetism. Naturally, the electronic structure of such a system depends strongly on the magnetic configurations for both bulk and surface.[1-4,6,7]

However, none of the EuM$_2$As$_2$ family members has been studied thoroughly. For example, both band calculations and experiments suggest EuCd$_2$As$_2$ is a topological semimetal.[1,3] There is accumulating evidence for the deviation of the semimetallic behavior for EuCd$_2$As$_2$ with a low carrier density.[8-10] The semiconducting nature with the bandgap around 0.77 eV has been observed for both bulk[8] and surface.[11] Earlier observation of the metallic transport in EuCd$_2$As$_2$ could result from band bending at the surface.[9] To investigate the electronic properties at the surface, scanning tunneling microscopy/spectroscopy (STM/S) is an ideal probe, which offers the direct measurement of the local density of states (LDOS) at the atomic level. STM experiments conducted on EuIn$_2$As$_2$ yielded a partial Eu-terminated surface after cleavage at room temperature[12] and 20 K.[13] However, such partial Eu termination was not observed in EuCd$_2$As$_2$.[11] Given that the crystal structure of EuCd$_2$As$_2$ (*P-3m*1) is slightly different from EuIn$_2$As$_2$ (*P*6$_3$/*mmc*), comparison of the surface termination between them may not be straightforward, however.

In this article, we report investigations of the surface electronic structures of EuZn$_2$As$_2$ single crystals with STM/S and first-principles calculations. We choose EuZn$_2$As$_2$ because it forms the same structure as EuCd$_2$As$_2$. Due to closer packing, Eu orders antiferromagnetically at $T_N$ ~ 19 K, double that for EuCd$_2$As$_2$.[14] Given its weaker spin-orbit coupling, a small bandgap is shown in the electronic structure, predicting EuZn$_2$As$_2$ a semiconductor with a small bandgap (~30 meV).[15] The surfaces were created by cleaving single crystals at liquid nitrogen temperature. Through STM/S, we find several types of surface and subsurface defects that have a profound impact on the surface electronic structure, thus allowing us to distinguish different terminations. There are distinct differences between the electronic structures of Eu and AsZn-terminated surfaces, which are elucidated by STS measurements. In particular, we observe (1) a large bandgap (~1.5 eV) at the pristine surface, (2) the gap size is reduced locally surrounding defects, and (3) the impact of defects is much more dramatic on the Eu surface than the AsZn surface.

**Results and Discussion**

**Influence of heterogeneities on the surface electronic structures of EuZn$_2$As$_2$**

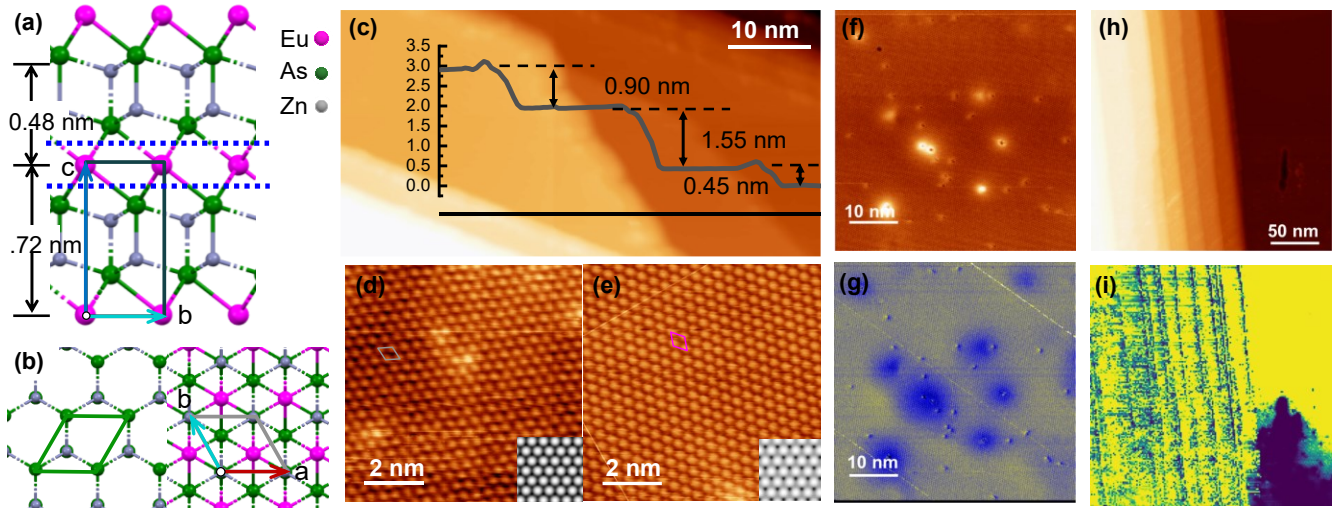

**Figure 1. Crystal structure and morphology of EuZn$_2$As$_2$.** **(a)** Sideview of the EuZn$_2$As$_2$ crystal structure. Two blue dashed lines represent possible terminations: Eu (magenta balls) and AsZn (green and grey balls). The black rectangle represents the unit cell in the *b* and *c* directions. **(b)** Top view of the AsZn and Eu terminations. The *ab*-plane unit cells are represented with solid green and gray parallelograms. **(c)** A large-scale STM topography of EuZn$_2$As$_2$ showing multiple steps and terraces with various step heights (V$_{bias}$ = -1.5 V, I$_t$ = 50 pA). The line profile along the black line is shown as an overlay. **(d)** and **(e)** STM images of the AsZn (d) and Eu (e) terminations with marked unit cells. V$_{bias}$ = -1.5 V, I$_t$ = 20 and 70 pA, respectively. Their simulated STM are presented as insets. **(f)** A typical high-resolution STM topography of the AsZn surface (V$_{bias}$ = -1.5 V, I$_t$ = 50 pA) with defects. **(g)** STS map collected simultaneously with (f). **(h)** An STM topography of EuZn$_2$As$_2$ with steps and terraces. **(i)** Integrated empty states image between 0 V and 1.5 V collected within the same area as (h).

EuZn$_2$As$_2$ crystallizes in a trigonal structure with the space group P-3*m*1 (No. 164) with lattice constants of $a = b = 0.42$ nm, and $c = 0.72$ nm.[14] Fig. 1(a) shows the ball and stick model of EuZn$_2$As$_2$ in the *bc* plane, showing that the AsZn slab is separated by Eu along the *c* direction. When a crystal is cleaved, the bonds between Eu (magenta) and As (green) atoms are expected to break, at the positions indicated by the two blue dashed lines in Fig. 1(a). This is because the Eu-As bonding (with a bond length of 0.31 nm) is weaker compared to the As-Zn bonding (with a bond length of 0.25 nm).[14] With two possible terminations shown in Fig. 1(b), Eu or AsZn, step heights of 0.72 nm, or 0.48 nm, or their multiples could be observed. If only one termination were present, the step height is expected to be 0.72 nm or its multiples. Fig. 1(c) presents an STM image of $EuZn_2As_2$ after an *in-situ* cleavage at 77 K. By scanning along the black line, we obtain the line profile as shown in Fig. 1(c). It is obvious that multiple terraces with different step heights are present. Identifying the terminations of these terraces is a nontrivial task, as both the AsZn and Eu terminations have the same lattice constant and symmetry as illustrated in Fig. 1(b). Figs. 1(d) and 1(e) show the STM images with atomic resolution. Through the presence of the unique defects combined with DFT calculations for the AsZn and Eu terminations, we identify that Fig. 1(d) is the AsZn surface while Fig. 1(e) is the Eu surface, with details described below.

As can be seen from Figs. 1(d) and 1(e), there are defects at the cleaved surfaces, such as steps generated during cleavage, surface imperfections including absorbates, substitute atoms, vacancies, and scan-induced damage. Fig. 1(f) shows a large-scale surface morphology of $EuZn_2As_2$ with atomic resolution, where defects are clearly seen. Fig. 1(g) is the STS map collected simultaneously with the STM image shown in Fig. 1(f). Note that atomic-level defects can drastically alter the surrounding electronic structure as indicated in blue "clouds" in the STS image with an affecting area of nearly 10 nm × 10 nm. To better illustrate the defects-induced surface electronic property changes, a much larger area of STM and STS images of $EuZn_2As_2$ are shown in Figs. 1(h) and 1(i), respectively. From these images, multiple step heights are clearly seen, reflecting the presence of both terminations. Note that blue contrast extends beyond step edges, spreading into terraces and appears around defects (on the right large terrace). In brief, the influence of imperfections on the surface overwhelms the electronic properties over a much larger area than the physical extent of the defects. As such, it is difficult to distinguish the two possible surface terminations using conventional STM and STS imaging techniques.

**Termination identification using substitutional defects**

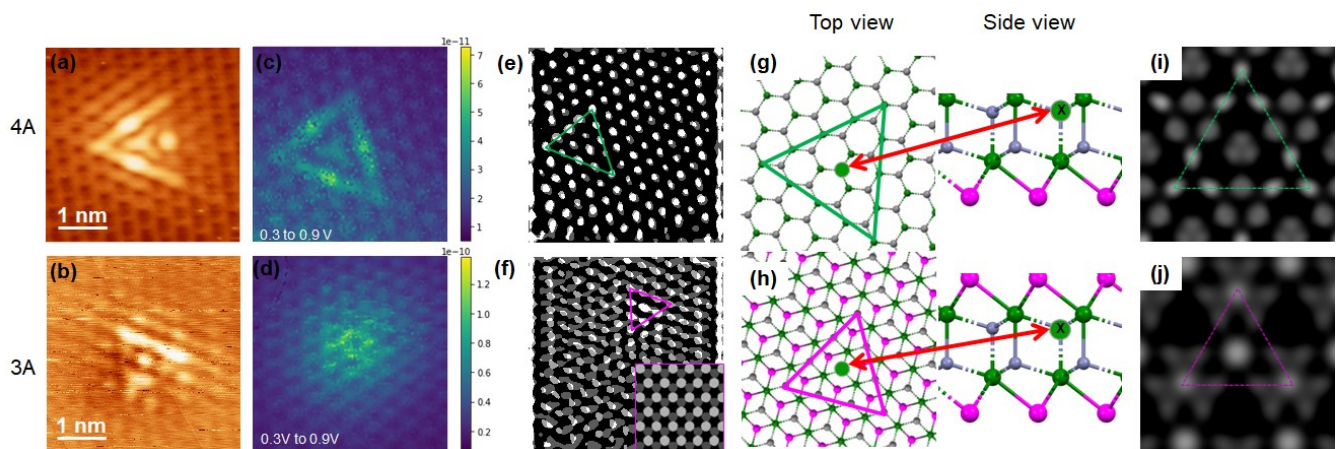

**Figure 2. STM images, electronic properties, and structure models of the substitutional defects.** STM images of **(a)** 4A and **(b)** 3A defects (set point: $V_{bias}$ = -1.5 V, $I_t$ = 50 pA). **(c-d)** Normalized integrated empty states from 0.3 V to 0.9 V of the 4A and 3A defects, respectively. Note the scale difference in the two-color bars. **(e-f)** Comparison of the superimpositions of two STS slices at 1.5 V and 1.075 V (50% transparency) from CITS maps, for 4A and 3A defects, respectively. The superposition of simulated STM images at 2 V and 1 V (65% transparency) of Eu termination is shown as an inset in (f). **(g-h)** Top and side view of the structure model of the 4A and 3A defects (see text for details). Red arrows point to the positions of the defected atoms (atoms with cross). **(i-j)** Simulated STM images for the corresponding 4A and 3A defects at 1 V, respectively.

It is observed that a special group of substitutional defects can be used to differentiate the terminations because of their exclusivity on terminations, although the ascription of the terminations still cannot be identified only with the experimental data. Different from other structural defects like step edges or surface adsorbates, this group of equilateral-triangular-shaped defects, shows a modification to the local density of empty states around the defect (Figs. 2 and 3). Using the morphology and spectroscopy observations of the substitutional defects as inputs, deep kernel learning (DKL) software[16,17] was used to survey the surface to establish statistics on the correlations between the defects and the local surface termination (i.e., particular kind of terrace) This DKL method enables much faster acquisition of structure-property correlations compared to the traditional operator-driven modes. Coexistence of 3A defects (equilateral-triangular-shaped defects with three-atomic lengths) and 4A defects (equilateral-triangular-shaped defects with four-atomic lengths) on the same surface termination was never observed. However, coexistence of 3A and 3A' defects (see below and Fig. 3) was observed on the same terrace, and 4A and 2A' defects (see below and Fig. 3) were observed exclusively on the other kind of terrace.

Figs. 2(a) and 2(b) present STM images with 4A and 3A defects, respectively. Besides their size differences, the sides of the 3A triangle are brighter at the corners while the sides of the 4A triangle feature a bright spot at their center. The 4A defect has a three-pointed star shaped weight center, while the weight center of 3A is a single bright point. The apparent heights of the 4A and 3A defects in the STM images are in the range between 5 and 20 pm, which is too small to be surface adsorbates. For the substitutional defects, the dI/dV maps are integrated over the bias range that shows visible contrast. The data is further normalized by the energy range from 0.3 V to 0.9 V and are shown in Figs. 2(c) and 2(d) for the 4A and 3A defects, respectively. The triangle shape for 4A remains unchanged through most of the energy slices (Fig. 2(c)), while the relative brightness of the 3A defect shifts at different energies resulting in the fuzzy integrated map (Fig. 2(d)).

Shown in Fig. 2(e) and 2(f) are the superimpositions of the STS slices at 1.5 V and 1.075 V (50% transparency) of the surfaces with 4A and 3A defects, respectively. Apparently, the STS maps indicate different bias dependence between 4A and 3A contained surfaces. As outlined with green (Fig. 2(e)) and magenta (Fig. 2(f)) triangles, the atoms for the two energy slices overlay on top of each other in the whole areas for the surface containing 4A defects. On the contrary, the atoms outside the triangle surprisingly shift at the two energies for the surface containing 3A defects, although the atoms within the magenta triangle overlay each other. This implies that, for the surface surrounding the 3A defects, the visible atoms in STS switch from one element to another between the bias of 1.5 V to 1.075 V, while those within the 3A defect area (magenta triangle) remain unchanged.

To understand the correlations between the defects and terminations, density functional theory (DFT) calculations are performed to sort out the observed distinct features of 4A and 3A defects. The simulated STM images of two pristine $EuZn_2As_2$ surfaces at -1.5 V are shown in the insets of Figs. 1(d) and 1(e), respectively. Although they look similar, the visible atoms on the AsZn termination are As atoms, while on the Eu termination, they are Zn atoms. Within the experimental bias range, the brightest contrast is always from the As atom for the AsZn termination. But for the Eu termination, the brightest contrast alters from Zn (< 0.5 V) to As (0.5 V – 1.5 V) to Eu (> 1.5 V). The density of states (DOS) switching between 1 V to 2 V is presented in the superimpositions of the two images shown in the inset of Fig. 2(f), where the bright atoms alternate from As (1 V) to Eu (2 V).

Combining the experiment observation and DFT calculations, we identify the surface with 4A defects as the AsZn termination and the one with 3A defects as the Eu termination.

A structure model with a substitutional atomic defect is proposed based on the STM/S signatures on 4A and 3A defects. The model proposes that the observed 4A and 3A defects have the same origin (i.e., an As atom substituting for Zn site), and the 4A and 3A signatures are from the projection of the same substitutional defect on the AsZn and Eu terminations, respectively. The size of the observed triangles is different due to the depth difference of the substitutional atom with respect to the surface. The crossed green atom in the center of the triangle in Figs. 2(g) and 2(h) is where the substitutional atom is located (Zn site), the side view shows the depth of the substitution (relative to the surfaces). The follow-up question is what atom occupies the Zn site? In $EuZn_2As_2$, As acts as an electron acceptor and it is unfavorable for either Eu or Zn to replace As. A more feasible scenario is that As is being misplaced, and ultimately acting as an electron donor as well. Considering the radius differences among $Eu^{2+}$ (1.25 Å), $As^{3+}$ (0.7 Å), and $Zn^{2+}$ (0.7 Å), it is likely that $As^{3+}$ can be kinetically misplaced into a $Zn^{2+}$ site such that As functions as an electron donor. This scenario is checked through DFT calculations. The simulated STM images based on the DFT calculations of the corresponding defects are shown in Figs. 2(i) and 2(j), respectively. Note that the simulated STM images reproduce the main features of the experimental results. DFT based simulations of SM images were also carried out for other substitutional possibilities, none of which resembled the experimental images.

We conducted similar experimental and DFT studies on the surface vacancies. Shown in Fig. 3 are two types of vacancy defects marked as 2A' and 3A', respectively. The 2A'-type vacancy is on the AsZn surface while the 3A'-type vacancy is on the Eu surface based on our assignments above, and the finding of 2A' defects only on the AsZn surface termination with 4A defects, and the finding of 3A' defects only on the Eu surface termination with 3A defects. The line profile across the 2A'-type vacancy gives a height decrease of ~70 pm, indicating a missing atom rather than a substitution. The image of 2A'-type vacancy is bias dependent: it shows a threefold symmetry petal at 1.5 V (see the insert of Fig. 3(a)) but a void at -1.5 V (see Fig. 3(a)). On the other hand, the 3A'-type vacancy changes from a bright triangle with a hollow center under negative bias (see Figs. 3(b) and 3(d)) to a uniform dark triangle when turning to a positive bias (Fig. 3(f)). Structural models of 2A' defects on the AsZn termination and 3A' defects on the Eu termination are presented in Figs. 3(g) and 3(h), respectively. Similar to the 4A and 3A defects, the 2A' and 3A'defects have same origin (Zn vacancy), and they are the projection of the Zn vacancy defect on the As-Zn and Eu terminations, respectively.

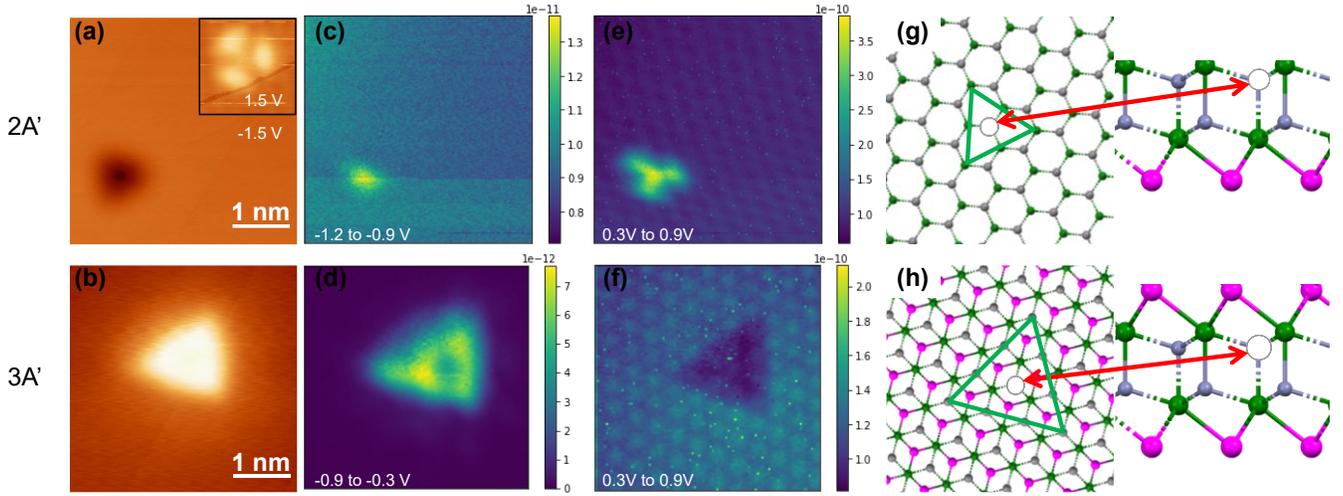

**Figure 3. STM images, electronic properties, and structure models of vacancy defects.** (a-b) STM images of **(a)** 2A' and **(b)** 3A' defects, set point: $V_{bias}$ = -1.5V, $I_t$ = 50pA. **(c-d)** Integrated occupied state maps from -1.2 V to -0.9 V for 2A' and from -0.9 V to -0.3 V for 3A' defects, respectively. **(e-f)** Integrated empty state maps from 0.9 V to 0.3 V for 2A' and 3A' defects, respectively. **(g-h)** Top and side view of the structure models of the vacancy defects (see text for details). Red arrows point to the positions of the defect location (white hollowed atom).

**Electronic properties of the defects and two terminations of EuZn$_2$As$_2$**

We have demonstrated that the existence of substitutional and vacancy defects greatly impacts the electronic properties of the surfaces. Fig. 4(a) displays the averaged dI/dV spectra collected through continuous imaging tunneling spectroscopy (CITS) at two pristine surfaces as well as defected areas on a linear scale. For easy comparison, we also replot the data on a log scale in Fig. 4(b). Remarkably, both the pristine Eu and AsZn surfaces reveal zero conductance between -1 V and +0.5 V in the conductance band. This clearly indicates that the pristine surfaces of EuZn$_2$As$_2$ are insulating with ~1.5 V bandgap. While structural defects on the surface extend the gap in the conductance band to above 0.6 V (yellow curve in Fig. 4(b)), the substitutional (3A and 4A) and vacancy (2A' and 3A') defects presented above considerably reduce the bandgap width. Especially, on the AsZn surface, the conductance band edge is lowered by the substitutional and vacancy defects 4A and 2A' to around 0.3 V (0.29 V for 2A', and 0.32 V for 4A), while the valence band edge remains unchanged. On the Eu surface, both the 3A- and 3A'-type defects drastically change the band gap widths through both valence and conductance band edges. For 3A (the black curve in Figs. 4(a)-(c)), the band gap is 0.9 V, with band edges as – 0.65 V and +0.28 V; for 3A' (the red curve in Fig. 4(a)-(c)), the band gap is 0.4 V, with band edges as – 0.34 V and +0.05 V.

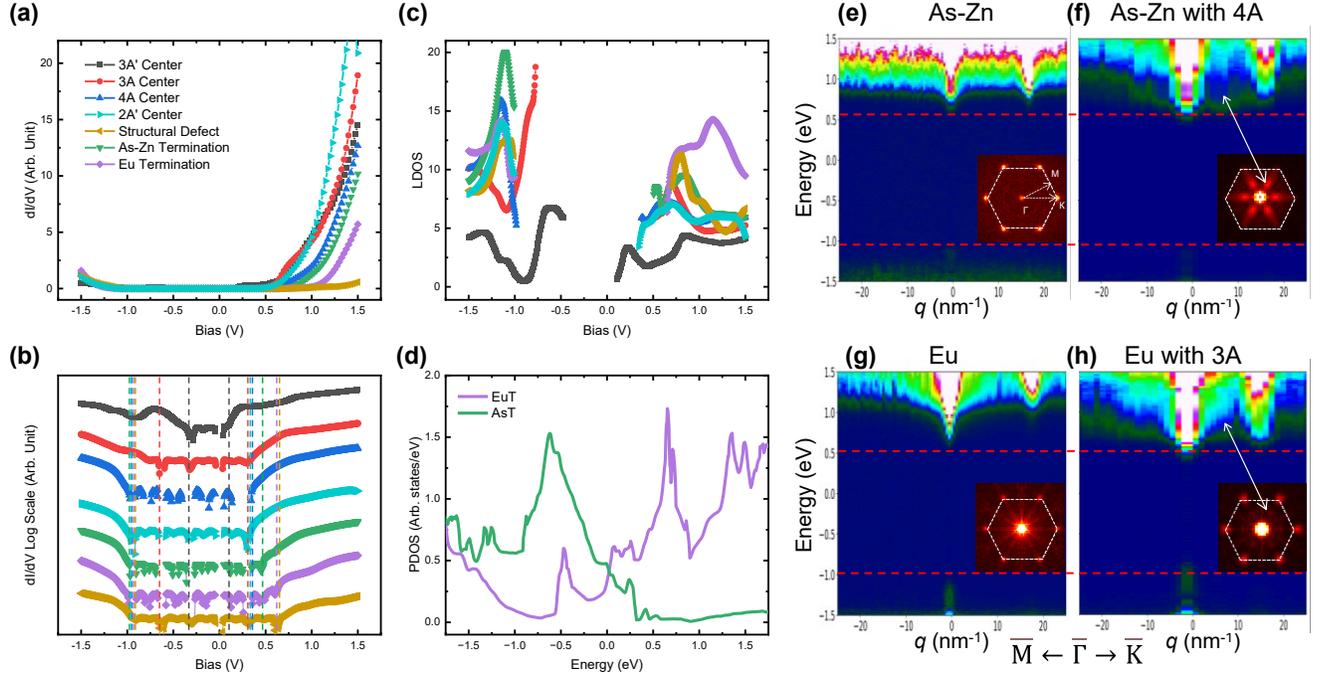

**Figure 4. Influence of heterogeneities on the surface electronic structures and QPI comparison of the two terminations and defects.** **(a)** Averaged STS spectra comparison of the two pristine surfaces, the two substitutional defects (3A and 4A), two vacancy defects (2A' and 3A'), and an area with structural defects. CITS set point: $V_{bias}$ = -1.5 V, $I_t$ = 200 pA. **(b)** STS spectra in (a) represented on a semi-log scale. Note the bandgap edges are marked by dashed lines. **(c)** Average local density of states (LDOS) derived from (a). **(d)** DFT calculated partial density of states (PDOS) of the two pristine terminations. (e-h) Energy vs. QPI signal in $\overline{K} - \overline{\Gamma} - \overline{M}$ directions for the pristine AsZn **(e)**, AsZn and with 4A defect **(f)**, pristine Eu **(g)**, Eu with 3A **(h)** defects, respectively. QPI maps from the FFT of the CITS are shown in the insets, energy at 1 V.

The bandgap width decreases by the presence of 4A and 3A defects are also shown in the quasiparticle interference (QPI) maps acquired from the fast Fourier transformation (FFT) of STS maps. The QPI analysis is conducted in the energy range between -1.5 V and +1.5 V, with the 25 mV energy intervals. The momentum maps are shown in Figs. 4 (e) and 4(g), with the QPI map obtained from the STS map at 0.6 V shown in the insets, respectively. The red dotted lines mark the valence and conductance band edges of the pristine surfaces. The penetration of the states into the gap is clearly shown in Figs. 4(f) and 4(h) for defected surfaces. The reciprocal lattice points of the atomic lattice are observable at $q_x \approx 17.4$ nm$^{-1}$, which is consistent with the atomic resolution STM images. QPI maps show similar dispersing signal at $\bar{\Gamma}$ for all cases, but additional QPI signal at $q \approx 8.7$ nm$^{-1}$ along the $\bar{\Gamma}$-$\bar{K}$ direction is found for the surface containing 4A defects, which corresponds to a new state inside the Brillouin zone. It occurs in the energy range of 0.5 ~ 1.5 eV as shown in Fig. 4(f), marked by white arrows, from the interference of scattered quasiparticles. This indicates that the surface-projected electronic structure strongly depends on the specific defects. We consider that defects affect the scattering potentials that influence the relative intensity between the scattering wavevectors, which even modify the surface band structure with new scattering wavevectors.

It is apparent that the decrease of the bandgap by substitutional and vacancy defects are in the compensation of the DOS away from the Fermi level. The LDOS, calculated from $(dI/dV)/(I/V)$ in Fig. 4(c), shows the comparison of the DOS changes of various cases. The two pristine terminations show highest filled state peak (AsZn, green, -1.1 V) and highest empty state peak (Eu, purple, +1.2 V), the trends consistent with the DFT calculated projected density of state (PDOS) shown in Fig. 4(d). The presence of the defects suppresses those peaks and pushes more states into the bandgap, forming in-gap states.

Quantitatively, the LDOS distribution for both the pristine AsZn and Eu surfaces is different from the calculated PDOS shown in Fig. 4(d). Our experiment observes a bandgap as large as ~1.5 eV, while calculations show conductive surfaces. The discrepancy may arise from two aspects. Our calculations were based on the A-type antiferromagnetic ground state at zero temperature for a slab with a clean and non-reconstructed surface. The experiment was carried out at 77 K, at which the system is non-metallic with complex magnetic interactions.[14,18] As introduced previously, the electronic properties of EuZn$_2$As$_2$ are extremely sensitive to magnetic interactions even at ambient field conditions. In our structural model, the unsaturated dangling bonds from a sharp cut of the surface tend to give in-gap surface states, yielding a conductive surface from calculation. Another possibility is the non-stichometry of experimental samples. According to experimental investigations, EuCd$_2$As$_2$ shows semimetallic behavior when the hole concentration is high[1,3] but is semiconducting when the hole concentration is low.[8-11] Bulk EuZn$_2$As$_2$ is also hole dominant[14] with nontrivial topological bands.[18] If the EuZn$_2$As$_2$ pristine surfaces preserve the same properties as that in the bulk, one expects that Zn vacancies introduce additional holes into the system. For As substitution into Zn sites, the situation is less clear. From the chemistry point of view (both ionic size and Zn function), As in the Zn site should behave as As$^{3+}$. The replacement of Zn$^{2+}$ by As$^{3+}$ will introduce an extra electron, thus reducing the overall hole concentration. Surprisingly, both substitutional and vacancy defects reduce the bandgap size regardless of the nature of the defects-induced carriers. This suggests defects-induced in-gap states as proposed for the surface of EuCd$_2$As$_2$.[9] What is more remarkable is that defects-induced bandgap modification is much more dramatic on the Eu surface than that on the AsZn surface: (1) the bandgap in the area around the Zn vacancy on the Eu surface is 0.4 V but 1.3 V on the AsZn surface, and (2) the gap measured in the As substituted area is 0.9 V on the Eu surface but 1.3 V on the AsZn surface. This implies that the Eu layer plays an extremely important role in the electronic properties of EuZn$_2$As$_2$, which is also consistent with the strong interplay between electronic structure and magnetism.

**Summary**

We have investigated the surface electronic properties of EuZn$_2$As$_2$ using STM/S. The surfaces were created by cleaving EuZn$_2$As$_2$ single crystals at 77 K and studied at the same temperature. Two surfaces, Eu- and AsZn-terminated, with identical surface structures, were identified through defects-induced LDOS changes, which are confirmed by DFT simulations. Several properties were observed: (1) both Eu and AsZn pristine surfaces are insulating with bandgap around 1.5 eV, (2) the surface bandgap is dramatically reduced due to defects located in the Zn site, either Zn vacancy or As substitution, and (3) defects-induced LDOS changes are more impactful on the Eu surface than on the AsZn surface. These experimental results obtained at 77 K cannot be explained by DFT calculations for T = 0 K, likely due to strong coupling of the electronic structure with magnetism. Our work sheds light on (1) the nature of the surfaces of EuZn$_2$As$_2$, (2) defects effects on the surface electronic properties, and (3) reconciliation of DFT simulations. Future work on the spin-polarized STM/S will help understand the role of magnetism and topology.

**Methods**

Single crystals of EuZn$_2$As$_2$ were grown via the flux method using Sn with details described elsewhere.[1] The crystal structure was determined via x-ray diffraction at room temperature, which belongs to *P*-3*m*1 space group (#164) with the lattice constants $a = b = 4.2093$ Å, and $c = 7.153$ Å. The magnetization measurements indicate the antiferromagnetic transition temperature $T_N = 19$ K. The in-plane electrical resistivity shows metallic behavior above ~ 200 K but nonmetallic below 200 K.[1]

The EuZn$_2$As$_2$ single crystals were cleaved on a cold stage which was cooled by liquid nitrogen under ultrahigh vacuum condition ($< 1\times10^{-10}$ torr). The samples were immediately *in situ* transferred into a pre-cooled homemade high magnetic field low temperature scanning tunneling microscope. STM/STS experiments were carried out at 77 K with base pressure lower than $1\times10^{-10}$ Torr using electrochemically etched Tungsten tips (W tip). All W tips were conditioned and checked using a clean Au (111) surface before each measurement. Topographic images were acquired in a constant current mode with a bias voltage applied to samples. All the spectroscopies were obtained using a lock-in amplifier with bias modulation $V_{rms} = 20$ mV at 977 Hz. Point spectroscopies, line spectroscopies, and Current-Imaging-Tunneling-Spectroscopy (CITS) were collected at particular single point, along a defined line, and over a grid of pixels at bias ranges around Fermi level using the same lock-in amplifier parameters, respectively.

The data analysis related to structure-property correlation was achieved using the deep kernel learning (DKL) framework. The DKL consists of a deep neural network (DNN) in combination with a Gaussian Process (GP) based regressor. The DNN consisted of three layers with the first, second, and third layers containing 64, 64, and 2 neurons respectively. The output of the last layer was treated as inputs to the GP regression. The inputs for the DKL were feature vectors extracted from the STM morphology image. This was trained against a suitable property scalar that was processed from the spectroscopic data. During the active learning process, the upper confidence bound acquisition function (UCB) was utilized to guide experimental exploration. In the UCB method, the β parameter was annealed in the range of 10 – 0.001, with a 10 % reduction in every successive iteration. Open-source python packages Atomai (https://github.com/pycroscopy/atomai) was used for image processing and feature extraction, while Gpax (https://github.com/ziatdinovmax/gpax) was used for the DKL-based training and GP regression. These were integrated with LabVIEW programs to access the STM controls.

The integration of the dI/dV hyperspectral image from CITS was carried out in custom python 3.9.12 code. At each topological point, the dI/dV signals were added together for all the bias slices within the bias range and then normalized based on the number of slices included. Then the matrix of the signal was utilized to construct a pseudo-color image. The upper limits of the images were manually selected to maximize the number of features shown in the image.

Density functional theory (DFT) calculations are performed using the projector augmented-wave (PAW) method as implemented in Vienna *ab initio* simulation package (VASP)[19,20]. The exchange correlation energy is described by the generalized gradient approximation (GGA) of Perdew-Burke-Ernzerhof (PBE) functional[21]. The valence configurations for each element are $5s^25p^66s^24f^7$ for Eu, $3d^{10}4s^2$ for Zn, and $4s^24p^3$ for As. To accurately describe the Coulomb correlations of the *f* orbitals, we use the spherically averaged DFT+U[22] with an on-site Coulomb interaction $U$ of 6 eV. The cutoff of kinetic energy for plane wave expansion is set to 520 eV and the atoms are relaxed until the Hellmann-Feynman forces on each atom is below 0.01 eV/Å. We start from the five-atom primitive cell of bulk $EuZn_2As_2$ [as illustrated in Fig. 1(a)], and double the cell size by expanding the lattice parameter $c$ by a factor of two, yielding a ten-atom unit cell for simulations. A Γ-centered **k**-point mesh of 7×7×2 is used to sample the Brillouin zone of this unit cell. Lattice constants of $EuZn_2As_2$ unit cell are optimized to be $a = b = 4.25$ Å, and $c = 14.47$ Å, in good agreement with the experimental values ($a = b = 4.2$ Å, and $c = 14.4$ Å).

The unit cell of $EuZn_2As_2$ contains two Eu atomic layers along the [001] direction, which are denoted as a "double layer" hereafter. The magnetic configuration considered in the calculations is type-A antiferromagnetic with interlayer spins aligned antiparallelly along the *c* axis and intralayer spins aligned parallelly. We study the (001) surface of $EuZn_2As_2$ and two surfaces can be distinguished: As- and Eu-terminated surfaces. By stacking *seven* unit cells of $EuZn_2As_2$ along the [001] direction, followed by removing the both top and bottom AsZn layers, and an inclusion of vacuum along [001] with a thickness of 15 Å, we create a slab with two Eu-terminated surfaces. By further removing the outer Eu layer from both the top and the bottom surface of the slab, while keeping the vacuum thickness the same, we create an As-terminated slab. We use the **k**-point grid of 7×7×1 for all pristine surface calculations in the described slab geometry. To incorporate the substitutional surface defects in the slab and to reduce the impacts of the finite-size effect, we create a 3×3×1 supercell for both terminations based on the previous slab geometry (576 atoms and 594 atoms for As- and Eu-terminated slabs, respectively) and introduce *one* defect on each surface of this 3×3×1 supercell. The calculations for these larger supercells are performed using a single **k**-point at Γ. STM simulations are visualized using the p4vasp software package[23].

**Data Availability**

The data sets that support the findings in this study are available from the corresponding author upon request.

**Code Availability**

The codes that support the findings in this study are available from the corresponding author upon request.

**Acknowledgments**

The STM work of this research was conducted at the Center for Nanophase Materials Sciences, which is a DOE Office of Science User Facility. Work at USC (R.J.) was partially supported by the grant No. DE-SC0024501 funded by the U.S. Department of Energy, Office of Science. S.M. was supported by the startup fund from the USC and an ASPIRE grant from the VPR's office of USC. This work used the Expanse supercomputer at the San Diego Supercomputer Center through allocation PHY230093 from the Advanced Cyberinfrastructure Coordination Ecosystem: Services & Support (ACCESS) program, which is supported by National Science Foundation Grants No. 2138259, No. 2138286, No. 2138307, No. 2137603, and No. 2138296.

**Author Contributions**

D.K., G.N. and Z.G. conducted the STM/S studies and data analysis, D.K. and Z.G. conducted the QPI analysis; S.K., and S.M. conducted the DFT calculation, P.R. and R.J. grew the crystals; R.J., and Z.G. designed the experiments; all authors contributed to the discussion and writing of the manuscript.

**Competing Interests**

The authors declare no competing interests.

This manuscript has been authored by UT-Battelle, LLC under Contract No. DE-AC05-00OR22725 with the U.S. Department of Energy. The United States Government retains and the publisher, by accepting the article for publication, acknowledges that the United States Government retains a non-exclusive, paid-up, irrevocable, world-wide license to publish or reproduce the published form of this manuscript, or allow others to do so, for United States Government purposes. The Department of Energy will provide public access to these results of federally sponsored research in accordance with the DOE Public Access Plan (http://energy.gov/downloads/doe-public-access-plan)